# Conditions for Relativistic Magnetic Reconnection under the Presence of Shear Flow and Guide Field

Sarah Peery, Yi-Hsin Liu, and Xiaocan Li
Dartmouth College, Hanover, NH 03750, USA

## Abstract

The scaling of the relativistic reconnection outflow speed is studied in the presence of both shear flows parallel to the reconnecting magnetic fields and guide fields pointing out of the reconnection plane. In nonrelativistic reconnection, super-Alfvénic shear flows have been found to suppress reconnection. We extend the analytical model of this phenomenon to the relativistic regime and find similar behavior, which is confirmed by particle-in-cell simulations. Unlike the nonrelativistic limit, the addition of a guide field lowers the in-plane Alfvén velocity, contributing to slower outflow jets and the more efficient suppression of reconnection in strongly magnetized plasmas.

*Unified Astronomy Thesaurus concepts:* Plasma astrophysics (1261); Plasma physics (2089)

## 1. Introduction

Magnetic reconnection is an energetic process ubiquitous to many plasma systems, wherein magnetic field line connectivity is rearranged, releasing magnetic energy into plasma kinetic and thermal energy. In particular, magnetic reconnection in the relativistic regime, where the magnetic energy density is larger than the plasma rest energy density, has been the subject of recent studies. These conditions are relevant to black hole accretion disks, pulsar wind nebulae, and active galactic nucleus jets (Lyubarsky & Kirk 2001; de Gouveia Dal Pino et al. 2010; Giannios 2013; Barniol Duran et al. 2017), where reconnection may be responsible for observations of nonthermal particle acceleration and magnetic dissipation (Drenkhahn & Spruit 2002; Kirk & Skjæraasen 2003; Zenitani et al. 2009; Cerutti et al. 2012; Hoshino & Lyubarsky 2012; Guo et al. 2015). Reconnection at these sites is rarely antiparallel and often accompanied by an out-of-plane guide field and and other asymmetries, such as background plasma shear flows (Zenitani et al. 2009; Mbarek et al. 2022). Thus, the dynamics over a range of parameters are of interest. In this work, we will focus on the effects of an out-of-plane guide field and sheared plasma flows parallel to the reconnecting component of the magnetic field, such as can be found in jets with helical magnetic fields (Coroniti 1990; Boccardi et al. 2016; Wang et al. 2021).

Relativistic shear flows have been previously investigated in some contexts (Sironi et al. 2021; Wang et al. 2021). However, the focus has been predominantly on plasma energization. In the nonrelativistic regime, the shear flow effects on reconnection dynamics are more well studied (La Haye et al. 2010; Nakamura et al. 2013; Mahapatra et al. 2021; Paul & Vaidya 2021). It has been shown that shear flows have a stabilizing effect on the tearing instability and can decrease the reconnection rate in the nonlinear stage. At super-Alfvénic flows, primary reconnection is suppressed altogether (La Belle-Hamer et al. 1994; Chen et al. 1997; Li & Ma 2010; Hosseinpour et al. 2018). At sub-Alfvénic speeds, Cassak (2011) presented an argument for the scaling of the reconnection rate and outflow velocity as they decrease with the shear flow. This work will extend a similar argument to the relativistic regime and consider additional effects due to the guide field.

We expect the largest impact on behavior to come from the dependence of the in-plane Alfvén velocity on the guide field in the relativistic limit; a limit that can be formally defined as where the magnetization parameter $\sigma \equiv B^2/h' \gg 1$ with enthalpy $h' = \rho' c^2 + [\gamma/(\gamma - 1)]P'$, where $\rho'$ is the mass density, $P'$ is the pressure in the fluid proper frame, and $\gamma$ is the ratio of specific heats. In relativistic plasmas, the total Alfvén velocity is $V_A/c = \sqrt{\sigma/(1 + \sigma)}$, which approaches the speed of light for large $\sigma$ (Kagan et al. 2015). The relevant quantity for reconnection is the projection of the total Alfvén speed into the outflow direction $V_{Ax} = c\sqrt{\sigma_{rec}/(1 + \sigma_{rec} + \sigma_g)}$, where $\sigma_{rec}$ and $\sigma_g$ are the $\sigma$ based on the reconnecting and guide-field components, respectively. It is clear that this relevant Alfvén speed can be decreased by the presence of a guide field (Liu et al. 2015), unlike in the nonrelativistic limit. Previous studies of relativistic reconnection, including some resistive MHD and kinetic simulations, suggest a fast normalized reconnection rate $\sim$0.1, even with pair plasma (Lyutikov & Uzdensky 2003; Lyubarsky 2005; Hesse & Zenitani 2007; Takahashi et al. 2011; Liu et al. 2015, 2020; Ripperda et al. 2019). Some conclude that reconnection is slowed as the background guide field is increased (Zenitani & Hesse 2008), but have found no cases where reconnection would be entirely suppressed. However, in combination with the presence of a shear flow, the guide field will have a large effect on determining the regime where reconnection can proceed.

In the first section of this paper, we derive from the relativistic magnetohydrodynamics (RMHD) equations a model for the scaling of the reconnection outflow as a function of shear flow and guide-field strength. We then perform 2D kinetic simulations to compare suppression behavior to the predictions for a range of guide fields. Last, we discuss the conclusions and implications based on our results.

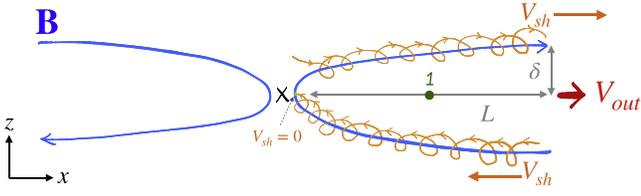

**Figure 1.** A diagram of the diffusion region. The blue lines show reconnected magnetic field lines projected on the reconnection plane. The orange arrows illustrate background plasmas streaming along the reconnected field line. The inflowing plasma at the lower right quadrant with speed $V_{\rm sh}$ toward the $x$-line deposits momentum on the reconnected field line to the right of the $x$-line, while the plasma at the upper right quadrant does not.

## 2. Scaling Theory

In this section, we present an analytical model that predicts the scaling behavior of relativistic reconnection in the presence of in-plane shear flows, based upon the intuition that both shear flows and the motional electric field in the presence of a guide field introduce momentum that must be overcome for reconnected field lines to be advected out of the diffusion region, allowing the reconnection process to continue. As illustrated in Figure 1, where the shear flow is aligned with the outflow jet direction, the shear flow does not affect the motion of the reconnected field line, because particles stream freely along the magnetic field lines toward the outflow region. Thus, the force balance considers only the sides where the shear flow and outflow are in opposing directions, and inflowing plasma is funneled toward the regions of high magnetic curvature. Using the standard special RMHD equations (Nakamura 2010), we derive the scaling of reconnection outflow speeds from the conservation of momentum.

We use 2D reconnection and look for a steady-state solution. We consider the case where the reconnecting field and velocity shear flow are both antisymmetric across the current sheet and include an out-of-plane guide field.

The RMHD equations are given as

$$\partial_\nu \left[ h' U^\mu U^\nu + P' \eta^{\mu\nu} + \frac{1}{4\pi}\left( F^\mu_\lambda F^{\nu\lambda} - \frac{1}{4}\eta^{\mu\nu} F^{\lambda\delta} F_{\lambda\delta} \right) \right] = 0, \quad (1)$$

$$\partial_\nu (\rho' U^\nu) = 0, \quad (2)$$

where $U^\mu = [\Gamma, \Gamma V_x, \Gamma V_y, \Gamma V_z]$ is the fluid four-velocity and the Lorentz factor is $\Gamma = 1/\sqrt{1-(V/c)^2}$. The Greek index runs from integers 0 to 3, denoting the time and spatial components. $\rho'$ is the plasma proper mass density, $P'$ is the proper pressure, $h' = \rho' c^2 + \gamma/(\gamma-1) P'$ is the enthalpy, $\eta^{\mu\nu} = \text{diag}(-1, 1, 1, 1)$ is the flat spacetime metric, $\partial_\nu$ is the four-derivative, and $F_{\lambda\delta}$ is the electromagnetic field tensor. In this paper, primed quantities denote proper quantities, and they include both electrons and positrons, since pair plasmas are expected to dominate the astrophysical systems of interest (e.g., Chen & Fiuza 2023).

Taking the spatial components of RMHD (i.e., $\nu = 1, 2, 3$ in Equation (1)), we recover the more familiar conservative form

of the momentum and continuity equations:

$$\partial_t \left( \frac{\Gamma^2 h'}{c^2} \mathbf{V} + \frac{\mathbf{E} \times \mathbf{B}}{4\pi c} \right)$$
$$+ \nabla \cdot \left[ \frac{\Gamma^2 h'}{c^2} \mathbf{V}\mathbf{V} + \left( P' + \frac{E^2}{8\pi} + \frac{B^2}{8\pi} \right) \mathbf{I} - \frac{\mathbf{E}\mathbf{E}}{4\pi} - \frac{\mathbf{B}\mathbf{B}}{4\pi} \right] = 0, \quad (3)$$

$$\partial_t (\Gamma \rho') + \nabla \cdot (\Gamma \rho' \mathbf{V}) = 0. \quad (4)$$

Considering the cold limit with $P' \simeq 0$, in the steady state, we can then combine these two equations into the force-balance equation:

$$\rho' \Gamma \mathbf{V} \cdot \nabla(\Gamma \mathbf{V}) + \nabla \cdot \left[ \left( \frac{E^2}{8\pi} + \frac{B^2}{8\pi} \right) \mathbf{I} - \frac{\mathbf{E}\mathbf{E}}{4\pi} - \frac{\mathbf{B}\mathbf{B}}{4\pi} \right] = 0. \quad (5)$$

We analyze the force balance in the $x$-direction at point 1 of Figure 1. Along the outflow symmetry line $B_x = E_x = 0$ and taking uniform $E_y$ and $B_y = B_g$ assumptions, we get

$$\rho' \partial_x \left( \frac{\Gamma^2 V_x^2}{2} \right) + \partial_x \left( \frac{E_z^2 + B_z^2}{8\pi} \right) - \frac{B_z \partial_z B_x}{4\pi} = 0, \quad (6)$$

which describes the balance between the plasma inertia, electromagnetic pressure gradient, and magnetic tension forces.

Plugging in gradient scales ($\partial_x \sim 1/L$ and $\partial_z \sim 1/\delta$) of the diffusion region and using $B_{z,\text{out}}/B_{x,\text{in}} \simeq \delta/L$ from $\nabla \cdot B = 0$, Equation (6) becomes

$$\sigma_{\text{rec}} \simeq \Gamma^2_{\text{out}}(V_{\text{out}}/c)^2 + \sigma_g (V_{\text{out}}/c)^2, \quad (7)$$

where $\sigma_{\text{rec}} \equiv B_{\text{in}}^2/(4\pi \rho' c^2)$ and $\sigma_g \equiv B_g^2/(4\pi \rho' c^2)$, and we have assumed $\rho'_1 = \rho_{\text{in}} = \rho$ and neglected $\partial_x(B_z^2/(8\pi))$ in the small $\delta/L$ limit. This simple relation determines the outflow speed without the shear flows.

Some limits yield familiar results. (1) Without a guide field, Equation (7) becomes $\sigma_{\text{rec}} = \Gamma^2(V_{\text{out}}/c)^2$, which can immediately be solved for $V_{\text{out}}/c = \sqrt{\sigma_{\text{rec}}/(1+\sigma_{\text{rec}})}$, recovering the relativistic Aflvén velocity. (2) In the strong-guide-field limit, we can solve for $V_{\text{out}}/c \simeq \sqrt{\sigma_{\text{rec}}/(1+\sigma_{\text{rec}}+\sigma_g)}$, if we take $V_{\text{out}}^4$ to be negligible, as expected (Liu et al. 2015).

Now, we consider the effect from shear flows by adding the force to Equation (6), which is needed to decelerate and stop the shear flow at the $x$-line—the ram pressure $\rho' \Gamma_{\text{sh}}^2 V_{\text{sh}}^2/(2L)$ and electric field pressure gradient $E_z^2/(8\pi L) \simeq (B_g^2/8\pi L)(V_{\text{sh}}/c)^2$ in discretized form. They can be viewed as the additional momentum carried by the background shear flow and Poynting flux (when $B_g \neq 0$).

The new relation that determines the outflow speed $V_{\text{out}}$ then becomes

$$\sigma_{\text{rec}} \simeq \Gamma_{\text{out}}^2 (V_{\text{out}}/c)^2 + \sigma_g (V_{\text{out}}/c)^2$$
$$+ \Gamma_{\text{sh}}^2 (V_{\text{sh}}/c)^2 + \sigma_g (V_{\text{sh}}/c)^2. \quad (8)$$

In the nonrelativistic limit ($\sigma_{\text{rec}} \ll 1$), it yields $\sigma_{\text{rec}} \simeq (V_{\text{out}}/c)^2 + (V_{\text{sh}}/c)^2$ or $V_{\text{out}} \sim \sqrt{V_A^2 - V_{\text{sh}}^2}$, recovering the expression of Cassak (2011). The outflow speed $V_{\text{out}}$ in the full Equation (8) can be evaluated numerically given the input parameters $\sigma_g$, $V_{\text{sh}}$, and $\sigma_{\text{rec}}$.

## 3. Simulation Setup

In the following, we validate our model against 2D particle-in-cell (PIC) simulations using the VPIC code (Bowers et al. 2009), which solves the fully relativistic dynamics of plasma and electromagnetic fields in flat space. The initial conditions are a force-free current sheet (Guo et al. 2014; Liu et al. 2020) with uniform background density and temperature in the proper frame. A density variation is implemented to satisfy the charge separation arising from the gradient of the motional electric field, generated by the background shear flow. We employ electron–positron pair plasma, with the particle mass of $m_i = m_e \equiv m$ and temperature ratio $T_p/T_e = 1$, motivated by studies that argue pair plasmas are relevant in astrophysical plasmas (Arons 2012; Barniol Duran et al. 2017). The density profile in the simulation frame is $n = \gamma_d n_0' + \delta n$, where the charge separation is added assuming $\delta n_i = -\delta n_e = \rho_c/2$ with the charge density $\rho_c$ found from Gauss' law. The relativistic factor of the current carrier is $\gamma_d = \sqrt{1 + J^2/(4n_0'^2 e^2 c^2)}$ with the current density $J$ found from Ampère's law.

The initial reconnecting field is $\mathbf{B} = B_{x0} \tanh(z/L)\hat{x} + B_y(z)\hat{y}$. An in-plane shear flow is implemented as $\mathbf{V} = V_{sh} \tanh(z/L_s)\hat{x}$, which is accompanied by the motional electric field $E_z = -V_x B_y/c$. If we require the guide field to have the asymptotic value $B_g$, the spatial dependence of $B_y(z)$ is found from the pressure balance $B_x^2 + B_y^2 - E_z^2 = B_{x0}^2 + B_g^2 - (V_{sh} B_g/c)^2$. To induce the reconnection x-line at the domain center, a magnetic perturbation of strength $\delta B_z = 0.13 B_{x0}$ is used in the initial condition.

In this paper, density is normalized to $n_0' = 1$, timescales are normalized to the electron plasma frequency $\omega_{pe} = \sqrt{4\pi n_0' e^2/m} = 1$, velocities are normalized to the speed of light $c$, and length scales are normalized to the ion inertial length $d_e = c/\omega_{pe}$. The boundary conditions are periodic in the x-direction, reflecting for particles and conducting for the fields in the z-direction. The system size is $L_x \times L_z = 384 d_e \times 384 d_e$, with $2048 \times 2048$ grid points and 100 macroparticles per cell.

The initial current sheet thickness is $L = 10 d_e$ and $L_s = 5 d_e$. The temperature in the bulk flow frame for both species is $T_0' = 0.5 mc^2$. For all runs, $\omega_{pe}/\Omega_{ce} = 0.066$, where $\Omega_{ce} = eB_0/mc$ is the ion cyclotron frequency. The asymptotic magnetization is then $\sigma_{x0} = B_{x0}^2/[2n_0' mc^2 + \gamma/(\gamma - 1)P_0'] = 70$ for the specific heat ratio $\gamma = 5/3$ and enthalpy $h_0' \simeq 3.25$.

## 4. Comparison

We ran an array of simulations with five values of the guide field: $B_g/B_{x0} = 0.01, 0.5, 1, 2,$ and $5$, with $V_{sh}$ varied from 0 to $0.9c$ for each. Figure 2 shows the field line structure in four runs with $B_g = B_{x0}$ and $V_{sh} = 0, 0.3c, 0.6c,$ and $0.9c$, respectively. To better visualize the signature of reconnection under background flows, we plot the in-plane magnetic flux transport (MFT) velocity (Liu et al. 2018; Li et al. 2021), defined as

$$\mathbf{U}_\psi = c \frac{E_y \hat{y} \times \mathbf{B}_p}{B_p^2}, \quad (9)$$

where $\mathbf{B}_p$ is the in-plane magnetic field. The MFT is useful here because it can be hard to distinguish reconnection jets from background shear flows using the plasma flows or the full $\mathbf{E} \times \mathbf{B}$ drift velocity.

The morphology in Figure 2 is representative of runs with a guide field $\gtrsim \mathcal{O}(B_{x0})$, which do not form plasmoids

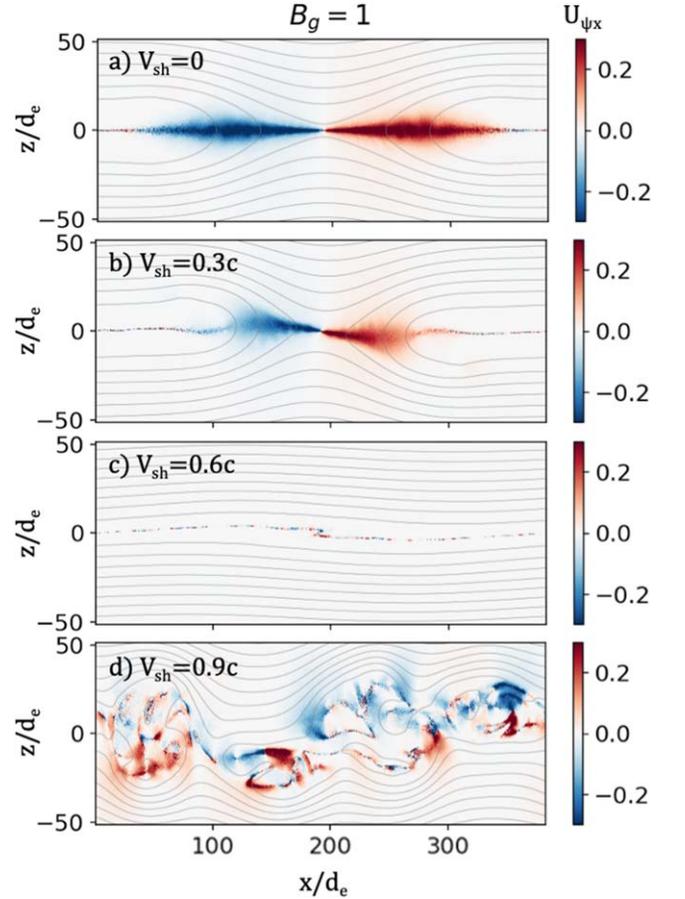

**Figure 2.** The magnitude of the MTF speed $U_{\psi x}$ with contours of the out-of-plane vector potential showing the in-plane magnetic field lines for simulations with $B_g = 1$ and $V_{sh} = 0, 0.3c, 0.6c,$ and $0.9c$.

(Liu et al. 2020). As the shear speed increases, the onset of reconnection is delayed, so to show the same evolutionary stage panel (a) is found at $7500/\Omega_{ce}$, while panels (b), (c), and (d) are found at $15000/\Omega_{ce}$. We clearly see the suppression of reconnection as the shear speed increases; with a sub-Alfvénic shear flow (Figure 2(b)), the reconnection onset is delayed and the jet has a lower outflow speed compared to the no-shear case (Figure 2(a)), but the reconnection is still fast. For $V_{sh}$ larger than the critical velocity (close to the in-plane Alfvén velocity $V_{Ax} \simeq 0.6c$; Figure 2(c)), reconnection does not start, consistent with its nonrelativistic counterpart (Cassak 2011). At a great enough shear flow (Figure 2(d)), vortices from the Kelvin–Helmholtz instability (KHI) start to form, and vortex-induced reconnection can occur (Nakamura et al. 2011; Hamlin & Newman 2013).

Because positrons and electrons behave similarly along the outflow, results will be shown only for the electrons. In the analysis of the simulation, all values are calculated as the average during the quasi-steady stage of reconnection, which we take to be when the reconnection rate $R > 0.7 R_{max}$. The reconnection rate is found as the rate of change in the flux converted between the primary x-point and o-point:

$$R \equiv \frac{cE_y|_{xline}}{B_{x0} V_{Ax0}} = \frac{1}{B_{x0} V_{Ax0}} \frac{d\Psi}{dt}, \quad (10)$$

where $\Psi = \max[A_y] - \min[A_y]$ along the $B_x = 0$ line, and $A_y$ is the out-of-plane component of the vector potential.

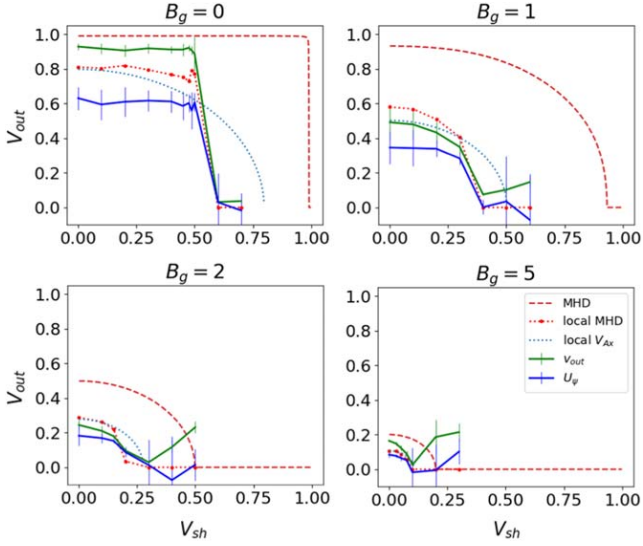

**Figure 3.** The simulation outflow velocity $V_{\rm out}$ and MFT speed $U_{\psi x}$ as a function of shear flow speed $V_{\rm sh}$, overlaid with predictions made from Equation (8), using local (labeled as "local MHD") and initial asymptotic parameters (labeled as "MHD"), and the local in-plane Alfvén velocity for comparison. The different panels have different guide fields.

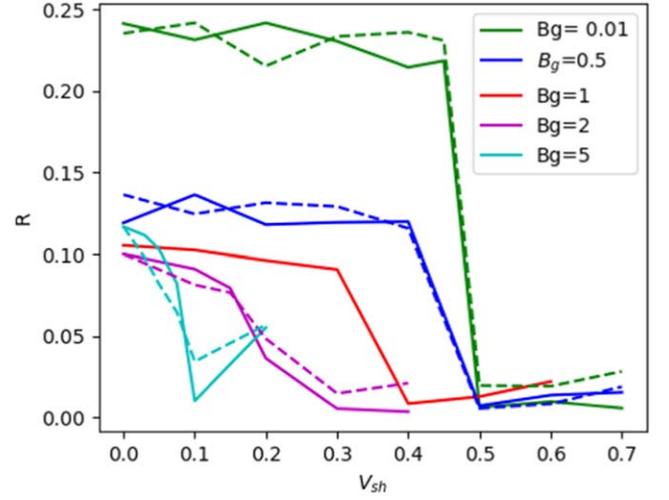

**Figure 4.** The average reconnection rate $R$ as a function of shear flow speed $V_{\rm sh}$ for five values of $B_g$. The dashed lines show the average outflow velocity $V_{\rm out}$, scaled for comparison of their trends (KHI runs are not shown).

Figure 3 shows five quantities plotted as a function of shear flow speed for different values of guide field; the simulation outflow velocity $V_{\rm out}$ and flux transport speed $U_{\psi x}$, predictions using Equation (8), and a circle of radius $V_{\rm Ax}$ (the in-plane Alfvén velocity). $V_{\rm out}$ is the maximum value of $V_x$ along the outflow symmetry line (where $B_x = 0$). For direct comparison, $U_{\psi x}$ is measured at the same position. In high-shear-flow cases with no dominant x-line, both $V_{\rm out}$ and $U_{\psi x}$ are primarily due to the KHI. The error bars are determined using the standard deviation during the quasi-steady period.

To evaluate Equation (8), we used two sets of input variables; the initial asymptotic conditions $\sigma_{\rm rec} = B_{x0}^2/4\pi h_0' \simeq 70$ and $\sigma_g = (B_g/B_{x0})^2 \sigma_{\rm rec}$ (dashed red line), and local values calculated from the simulations during reconnection (dotted red line). In the latter, we measured these values from the inflow edge of the diffusion region, defined as where the nonideal electric field $E_y' \equiv E_y + (\mathbf{V} \times \mathbf{B})_y/c$ falls to below 0.2 of its maximum value, for a cut across the center of the primary x-point. $h'$ is found from the local proper pressure tensor and density. We use the heat capacity ratio $\gamma = 5/3$ here, because the plasma remains mildly relativistic, as in $\sigma_{x0}$. The true $\gamma$ value may be in between this and $\gamma = 4/3$ for ultrarelativistic plasma, but the numerical calculation of $V_{\rm out}$ is not sensitive to this choice.

$B_{x,{\rm in}}$ is measured at the upstream edge. $h'$ is averaged across the current sheet; this takes into account the total enthalpy of the plasma being moved through the diffusion region and the drop of $\sigma_{\rm rec}$ due to the reduction of the reconnecting magnetic field in the large opening angle limit (Liu et al. 2017). It also allows for variations in the input parameters for cases with different shear flows, giving a much more accurate accounting of the momentum in the diffusion region. The circles of $V_{\rm Ax}$ in Figure 3 are calculated from the local magnetization in the cases with no shear flow present.

As we might expect, $V_{\rm out}$ is near the in-plane Alfvén velocity when there is no shear flow. It decreases slightly in the presence of slow shear flows, and then sharply near a critical cutoff velocity ($V_{\rm sh,c}$), which is also close to the in-plane Alfvén velocity. Largely, the simulation behavior mirrors the trends predicted in Equation (8), confirming that outflow jets of relativistic reconnection are suppressed in a manner similar to those observed in the nonrelativistic regime (Cassak 2011). However, in the relativistic case, the addition of a guide field reduces both the outflow velocity in the $V_{\rm sh} = 0$ case and the suppression velocity, consistent with the decrease in $V_{\rm Ax}$.

We do note that in the low-guide-field cases, the simulation results disagree slightly with the prediction. A similar trend in $V_{\rm sh,c}$ was observed in Sweet–Parker simulations (Cassak & Otto 2011), but it is hard to say if this has the same source or not. It is more likely that in runs that produce plasmoids, the spatial variation makes measurements (of $V_{\rm out}$ and $\sigma_{\rm rec}$) trickier and less comparable to the simple MHD predictions (dashed red line). In addition, energy conversion to other plasma channels, such as enthalpy flux, which becomes more important in a low guide field (Zenitani et al. 2009), is overlooked. We must also emphasize that the shear flow delays the onset of reconnection significantly as it approaches the critical value, which could also contribute to underestimating $V_{\rm sh,c}$ in our plots (e.g., the simulation with $V_{\rm sh} = 0.055c$ for $B_g = 0$ was allowed to run for up to 10 times as long and did not start reconnecting). Nevertheless, the simple model of Equation (8) is able to recover the overall behavior.

While no quantitative argument is made for it, the outflow velocity is closely related to the behavior of the reconnection rate; this is shown in Figure 4, where scaled outflow velocities (dashed lines) are plotted next to the average quasi-steady reconnection rate (solid lines) as a function of $V_{\rm sh}$. The similarity in their trends is unsurprising, given that the reconnection electric field can be approximated from $cE_y = B_{\rm rec}V_{\rm in} \simeq B_{\rm rec}V_{\rm out}(\delta/L)$. Or $R \propto V_{\rm out}$, if the aspect ratio is constant. This gives us confidence that Equation (8) and $V_{\rm out}$ may be used to examine the dynamics of the system as a whole. The upticks in outflow velocity at high-shear-flow values (the $B_g = 2$ and $B_g = 5$ cases), seen in Figure 3, are due to the formation of large-scale KHIs that cause fluctuations in $A_y$ and $V_x$ and can induce reconnection (Nakamura et al. 2011). This is not further explored in this study, so for clarity, cases with $V_{\rm sh} > V_{\rm sh,c}$ are not shown in Figure 4.

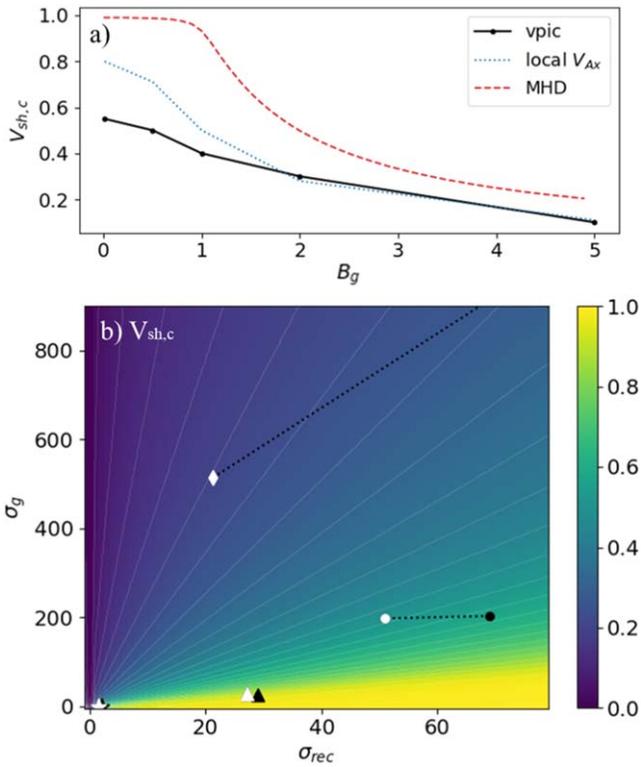

**Figure 5.** (a) The critical values of shear flow $V_{sh,c}$ needed to suppress reconnection, as a function of $B_g$. The dashed red line shows Equation (8) with initial conditions ($\sigma_{rec} = 70$, labeled as "MHD"). The black line shows the simulation values and the dotted blue line shows the in-plane Alfvén speed. (b) Predictions for the critical shear flow $V_{sh,c}$ from Equation (8) for a range of $\sigma_{rec}$ and $\sigma_g$. Contours are added for clarity. The markers represent the local parameters of the simulations, with $V_{sh} = 0$ (black) and $V_{sh}$ just below $V_{sh,c}$ (white). $B_g = 0$—star; $B_g = 0.5$—triangle; $B_g = 1$—dot; and $B_g = 2$—diamond. Higher $\sigma_g$ values are omitted for clarity.

The force-balance relation Equation (8) can be used to identify at what speed shear flows are capable of suppressing reconnection. Figure 5(a) shows critical values of shear flow ($V_{sh,c}$) as a function of the guide field for all values plotted in Figure 3: the MHD solution from Equation (8) with the initial asymptotic parameters, and the in-plane Alfvén velocity from the $V_{sh} = 0$ cases. In Figure 5(b), $V_{sh,c}$ is calculated for a range of magnetizations. Contours are shown for clarity, and to give some context, the local simulation parameters are marked for the $V_{sh} = 0$ case (in black) and the case with $V_{sh}$ just above $V_{sh,c}$ (in white). The $B_g = 5$ case is not shown, as it has $\sigma_g \sim 5000$ ($\sigma_{rec} \sim 60$).

These plots give us some insight into the interplay between guide field and shear flow. They can also be used to give us some prediction of where reconnection proceeds. The in-plane Alfvén velocity and measurements of $V_{sh,c}$ all decrease with the guide field, as expected. Local $\sigma_{rec}$ and $\sigma_g$ depend strongly on the initial guide field, as shown by the markers in Figure 5(b). In the $B_g = 0$ case, the plasma is more compressible (Zenitani & Hesse 2008) and can collapse to a very thin, high-pressure current sheet ($\sim d_e$ or less), surrounded by plasma that has a depleted magnetic field, leading to a local $\sigma_{rec} \sim 1$ (stars); whereas in the high-guide-field case (diamonds), a broad, low enthalpy current sheet leads to a local $\sigma_{rec} \sim 100$ (not shown). Interestingly, the local $\sigma_{rec}$ and $\sigma_g$ are found to vary with shear flow as well. At higher shear flow values, local $B_{rec}$ and $B_g$ decrease and $h'$ increases. The shear flow not only suppresses the formation of an outflow jet, but it also has an effect on the diffusion region thickness, as in Cassak (2011).

It's worth noting that even in low-guide-field cases, the simulation $V_{sh,c}$ is relatively low ($\Gamma = 1$) and approaches zero asymptotically with the $B_g$ strength. Therefore, while the presence of a strong guide field is not considered capable of suppressing reconnection, there will be some systems where the introduction of even a very small velocity shear will stop reconnection.

## 5. Summary

We have examined the scaling of collisionless relativistic reconnection in pair plasmas, under the presence of an in-plane antiparallel shear flow and an out-of-plane guide field. Introducing terms that accounted for the momentum associated with the shear flow to the force balance in the diffusion region, we solved for the outflow velocity as a function of guide-field strength and shear flow speed. Unlike its nonrelativistic counterpart, the electric field plays a large role here. One thus needs to include the momentum from the motional electric field $E_z = B_g V_{sh}/c$, which can also oppose the formation of the outflow jet. This MHD model, Equation (8), was validated using kinetic PIC simulations.

Although other effects on the diffusion region morphology can be identified, as in Cassak (2011), we leave those for the next article. Here, we have focused only on the suppression of $V_{out}$. A comparison of $V_{out}$ and the average reconnection rate in our simulations confirmed that $V_{out}$ is representative of general system behavior. With no shear flows present, the simulation $V_{out}$ is near the in-plane Alfvén velocity $V_{Ax} = c\sqrt{\sigma_{rec}/(1 + \sigma_{rec} + \sigma_g)}$ and thus decreases as guide-field strength increases. When a shear flow is introduced, the onset of reconnection is delayed, and the outflow speed decreases. We also found that the local $\sigma_{rec}$, the reconnecting field, decreases with $V_{sh}$. The velocity at which reconnection is suppressed completely is found to be near or below $V_{Ax}$ and thus is also strongly related to the guide-field strength.

Equation (8) can be used to examine where reconnection can proceed for a wide range of magnetization parameters. Notably, for large guide fields, $V_{sh,c}$ can be very small. This is valuable in studies relating to relativistic reconnection as a mechanism for observed particle acceleration. Specifically, the spine-sheath model for blazar jets, which has been used to explain gamma-ray observations, proposes sheared flows at the interface (Liang et al. 2013; Sikora et al. 2016). Reconnection in these conditions has already been found to produce nonthermal particle spectra (Sironi et al. 2021). Including asymmetries (Mbarek et al. 2022) and a more complete fluid closure in our model would be valuable in future work.

### ORCID iDs

Sarah Peery ● https://orcid.org/0009-0008-2916-4881
Yi-Hsin Liu ● https://orcid.org/0000-0001-5880-2645
Xiaocan Li ● https://orcid.org/0000-0001-5278-8029